\documentstyle[times,pramana,epsf,floats]{ias}
\begin{document}
\mark{{Quantum contact interactions}{Taksu Cheon}}
\title{Quantum contact interactions}

\author{Taksu Cheon}
\address{Laboratory of Physics, Kochi University of Technology,
Tosa Yamada, Kochi 780-8502, Japan\\
{\it taksu.cheon@kochi-tech.ac.jp} }
\abstract{
The existence of several exotic phenomena, such as duality and
 spectral anholonomy is pointed out 
in one-dimensional quantum wire with a single defect.
The topological structure in the spectral space which is behind these
phenomena is identified. 
}
\maketitle
%
\section{Introduction}
Since the successful applications of the quantum field theories
to the high energy particle physics, the low-energy 
phenomena  described by  the non-relativistic quantum mechanics 
has been regarded, in a way, as an area of rear guard action.
With the advent
of quantum information theory, however, it is recognized that 
a seemingly simple system in elementary quantum mechanical setting
can have highly nontrivial properties with potential technological ramifications. 

In this article, we point out a different kind of nontriviality of generic
low energy quantum mechanics other than that related to the entanglement.
The key concept here is the contact, or point interaction. 
Let us suppose that we have a one-dimensional quantum particle
subjected to a potential of finite support.  
If the range of the potential is small enough compared to the wave length of the
particle, one should be able to approximate the action of the potential as operating
at a single location.  In other word, one can regard the system as being 
free everywhere except the vicinity of of a single point. 
Every student of elementary quantum mechanics learns that such system is
described by a singular object called  Dirac's $\delta$-function potential, 
which induces the discontinuity of the space derivative of the wave functions. 
However, a natural question might arise to every naive mind: 
Why is the discontinuity 
allowed only for the derivative, not the wave function itself?
The answer to this question is not to be found in any elementary 
textbooks.  Indeed, it turns out that there is no good reason to reject
the discontinuity of the wave function itself.
We shall see in the followings that 
this possibility opens up a whole new vista to the problem. 

%
\section{Generalized point interaction described by U(2)}
We place a quantum particle on a one-dimensional line with a defect
located at $x = 0$.
In formal language, the system is described by the Hamiltonian
\begin{eqnarray}
\label{F1}
H = -{{1}\over{2}}{{d^2}\over{d x^2}} ,
\end{eqnarray}
defined on proper domains in the Hilbert space
${\cal H} = L^2({\bf R}\setminus\{0\})$.
We ask what
the most general condition at $x=0$ is.
%
We define the two-component vectors \cite{FT00},
\begin{eqnarray} 
\label{F2} 
\Phi =
  \left( {\matrix{ {\varphi (0_+)}\cr
                             {\varphi (0_-)} \cr }
            } 
  \right)
\ \ \
{\rm and} 
\ \ \
\Phi' =
  \left( {\matrix{{ \varphi' (0_+)}\cr
                  {-\varphi' (0_-)}\cr}
         } 
  \right)
\end{eqnarray}
from the values and derivatives of a wave function
$\varphi(x)$ at the left $x = 0_-$ 
and the right $x = 0_+$ of the missing point.
The requirement of self-adjointness of the Hamiltonian
operator (\ref{F1}) is satisfied if
probability current 
$j(x) = - i (
(\varphi^*)'\varphi - \varphi^* \varphi' )/2$
is continuous at $x = 0$. 
In terms of $\Phi$ 
and $\Phi'$, this requirement is expressed as
\begin{eqnarray} 
\label{F3}
\Phi'^\dagger \Phi - \Phi^\dagger \Phi' = 0 ,
\end{eqnarray}
which is equivalent to
$|\Phi-i L_0 \Phi'|$  $=$ $|\Phi+i L_0 \Phi'|$
with $L_0$ being an arbitrary constant in the
unit of length.  
This means that, with a two-by-two unitary matrix
$U\in U(2)$, we have the relation,
\begin{eqnarray} 
\label{F4} 
(U-I)\Phi+iL_0(U+I)\Phi'=0
\ .
\end{eqnarray}
This shows that the entire family $\Omega$ of
contact interactions admitted in quantum mechanics
is given by the group $U(2)$.
A standard parametrization for $U \in U(2)$ is
\begin{eqnarray} 
\label{F5} 
U =
e^{i\xi }
 \left( {\matrix{ {\alpha}    &{\beta}      \cr
                            {-\beta ^*}&{\alpha ^*}\cr }
           } \right) ,
\ \ 
%
\ \xi \in [0,\pi), \ \
\ \alpha,\beta \in {\bf C} ; \ \ 
| \alpha |^2+| \beta |^2 = 1 .
\end{eqnarray}
In mathematical term, the domain in which the Hamiltonian $H$ becomes self-adjoint
is parametrized by $U(2)$ --- there is a one-to-one correspondence between
a physically distinct contact interaction and a self-adjoint Hamiltonian \cite{SE86}.
We use the notation $H_U$ for the Hamiltonian with the contact interaction specified 
by $U \in \Omega$ $\simeq U(2)$.

If one asume $\Re\beta \ne 0$ and $\Im\beta\ne 0$, one can easily show that
(\ref{F2}), (\ref{F4}), (\ref{F5}) is rearranged in the form
\begin{eqnarray} 
\label{F6} 
 \left( {\matrix{ {\varphi(0_+)}\cr
                            {\varphi'(0_+)}\cr }
           } \right) 
=
\Lambda
\left( {\matrix{ {\varphi(0_-)}\cr
                            {\varphi'(0_-)}\cr }
           } \right) ,
\end{eqnarray}
with the form
\begin{eqnarray} 
\label{F7} 
\Lambda =
e^{i\lambda }
 \left( {\matrix{ {s}&{u}\cr
                            {v}&{t}\cr }
           } \right) ,
\ \ 
\ \lambda \in [0,\pi),
\ \
\ s,t,u,v \in {\bf R} ; \ 
st-uv = 1 .
\end{eqnarray}
This is the transfer matrix representation \cite{AG88}, which 
has been the treated as the standard form of generalized point interaction.  
But it is now obvious that, unlike the $U(2)$ representation, (\ref{F5}), 
the form (\ref{F7}) does not cover the whole family of 
generalized point interaction, thus does not gives complete parametrization.

%
\section{Fermion-boson duality}
The transfer matrix form (\ref{F7}) is non-the-less useful in making contact with
our intuition to the point interaction.
If we set $s=t=1$,  $uv=0$ has to be satisfied.  
By further choosing $\lambda=0$, one obtain two sets of  one parameter family of
transfer matrices
\begin{eqnarray}
\label{V1}
\Lambda_{\delta}(v) = 
\left(
\begin{array}{cc}
1 & 0 \\
v & 1 
\end{array} 
\right) ,
\ \ \ 
\Lambda_{\varepsilon}(u) = 
\left(
\begin{array}{cc}
1 & u  \\
0 & 1 
\end{array}
\right) .
\end{eqnarray}
The first one keeps the wave functions at $x=0_+$ and $0_-$ the same, while
giving the jump at $x=0$ for the value of their derivatives: This
clearly corresponds to the $\delta$ potential of strength $v$.
%
The second one gives the jump in the wave function 
itself at the location of the defect $x=0$.
We call this contact interaction as $\varepsilon$ potential with the strength $u$.  
It can be proven with
elementary algebra that this set of connection conditions is realizable as a 
singular zero-range limit of three-peaked structure \cite{CS98,EN01}.
It is anticipated from the construction that $\delta$ and $\varepsilon$ potentials play
a complimentary role.  An evident is that $\delta$ interaction at the origin 
has no effect on odd-parity states, 
while  $\varepsilon$ has no effect on even-parity states.  
A more quantitative expression of the complimentarity is obtained by 
considering the scattering properties of the $\delta$ 
and $\varepsilon$ potentials. 
We start by putting a generalized contact interaction at the origin 
on $x$-axis. 
Incident and outgoing waves can be written as 
\begin{eqnarray}
\label{eq3:1}
\varphi_{in}(x) & = & A(k) e^{ikx} + B(k) e^{-ikx} 
\ \ \ (x < 0), \\
\label{eq3:2}
\varphi_{out}(x) &  = & e^{ikx}
\ \ \ \ \ \ \ \ \ \ \ \ \
\ \ \ \ \ \ \ \ \ \ \ \ \
\ \ \ \ \ \ \ \
 (x > 0).  
\end{eqnarray}
The connection condition (\ref{F6}) is written as  
\begin{eqnarray}
\label{eq3:3}
\left( \! \!
\begin{array}{c}
1 \\
ik 
\end{array}
\! \! \right)
=
\Lambda 
\left(
\begin{array}{cc}
1 & 1 \\
ik  &  -ik 
\end{array}
\right)
\left( \! \!
\begin{array}{c}
A(k) \\
B(k) 
\end{array}
\! \! \right) .
\end{eqnarray}
The transmission and reflection coefficients are calculated respectively as
\begin{eqnarray}
\label{eq3:5}
T(k)=\left| \frac{1}{A(k)} \right|^2, \hspace{5ex}
R(k)=\left| \frac{B(k)}{A(k)} \right|^2 .
\end{eqnarray}
In case of $\Lambda=\Lambda_{\delta}(v)$, 
we obtain the well-known results  
\begin{eqnarray}
\label{eq3:6}
T_{\delta}(k) = \frac{k^2}{k^2+(v/2)^2}, \hspace{2ex}
R_{\delta}(k) = \frac{(v/2)^2}{k^2+(v/2)^2}. 
\end{eqnarray}
For $\Lambda=\Lambda_{\varepsilon}(u)$, we obtain 
\begin{eqnarray}
\label{eq3:7}
T_{\varepsilon}(k) = \frac{(2/u)^2}{k^2+(2/u)^2}, \hspace{2ex}
R_{\varepsilon}(k) = \frac{k^2}{k^2+(2/u)^2} .
\end{eqnarray}
One can observe that if $u=v$,  
$T_{\delta}(k)=T_{\varepsilon}(1/k)$ and 
$R_{\delta}(k)=R_{\varepsilon}(1/k)$ are satisfied. 
This implies that the low (high) energy dynamics of 
$\varepsilon$ potential is described by the high (low) 
energy dynamics of $\delta$ potential. 

The dual role of $\delta$ and $\varepsilon$ potentials becomes
more manifest 
when we consider the scattering of two {\it identical}
particles.  We now regard the variable $x$ as the relative 
coordinate of two identical particles whose statistics 
is either fermionic or bosonic.  The incoming and outgoing waves
are now related by the exchange symmetry.  We assume the form
\begin{eqnarray}
\label{eq3:8}
\varphi_{in}(x) & = &  e^{ikx} + C(k) e^{-ikx}
\ \ \ \ \ \ \ \ \ \ \ \ \
(x < 0) ,\\
\label{eq3:9}
\varphi_{out}(x) &  = & \pm e^{-ikx} \pm C(k) e^{ikx}
\ \ \ \ \ \ \ \ \ \
(x > 0) .
\end{eqnarray}
where the composite signs take $+$ for bosons and $-$ for fermions.
Important fact to note is that the symmetry (anisymmetry) of
$\varphi(x)$ leads to the antisymmetry (symmetry) of it's 
derivative $\varphi'(x)$.
The coefficient $C(k)$ becomes the scattering matrix.
The connection condition in (\ref{F6}) now reads
\begin{eqnarray}
\label{eq3:10}
\left[
  \left(
    \begin{array}{cc}
       1 &  1 \\
     -ik &  ik 
     \end{array}
    \right)
  \mp
    \Lambda 
   \left(
     \begin{array}{cc}
       1 &  1 \\
       ik &  -ik
      \end{array}
   \right)
\right]
\left( \! \!
\begin{array}{c}
C(k) \\
1 
\end{array}
\! \! \right)
= 0 .
\end{eqnarray}
We first consider the case for $\delta$ potential 
$\Lambda$ $ = \Lambda_\delta(v)$.
We obtain 
\begin{eqnarray}
\label{eq3:11}
C_\delta(k) &=& 1   \ \ \ \ \ \ \ \ \ \ \ \ \ \ \ \ \
  {\rm for \ fermions}, \\
C_\delta(k) &=& {{2ik + v} \over {2ik - v}} \ \ \ \ \ 
  {\rm for \ bosons}.
\end{eqnarray}
The first equation means that the $\delta$ function is
inoperative as the two-body interaction between identical
bosons, which is an obvious fact pointed out earlier.
Next we consider the case of $\varepsilon$ potential 
$\Lambda$ $ = \Lambda_\varepsilon(u)$. 
We have
\begin{eqnarray}
\label{eq3:12}
C_\varepsilon(k) &=& {{2ik + 4/u} \over {2ik - 4/u}} \ \ \ 
  {\rm for \ fermions}, \\
C_\varepsilon(k) &=& 1 \ \ \ \ \ \ \ \ \ \ \ \ \ \ \ \ \ \
  {\rm for \ bosons}.
\end{eqnarray}
One finds that the role of fermion and boson cases are exchanged:
The $\varepsilon$ potential as the two-body interaction  has 
no effect on identical bosons,
but {\it does} have an effect on the fermions.
Moreover, the scattering amplitude of fermions with
$\Lambda_\varepsilon(u)$ is exactly the same as 
that of bosons with $\Lambda_\delta(v)$ if the two
coupling constants are related as
\begin{eqnarray}
\label{eq3:13}
vu = 4.
\end{eqnarray}
Therefore, a two-fermion system with $\varepsilon$ potential 
is dual to a two-boson system with $\delta$ potential with role of the
strong and week coupling reversed.  
As expected, a natural generalization to 
$N$-particle systems exists \cite{CS99}.

%
\section{Spectral space decomposition and spiral anholonomy}
We now go back to the general $U(2)$ representation of the contact interaction,
and look at the structure of the parameter space more closely. 
Let us consider following {\it generalized parity} 
transformations \cite{TF00,CF01}:
\begin{eqnarray} 
\label{P1}
{\cal P}_1&:& \varphi (x) 
\longrightarrow  
({\cal P}_1\varphi)(x) 
= \varphi( - x),
\\
{\cal P}_2&:& \varphi (x) 
\longrightarrow  
({\cal P}_2\varphi)(x) 
= i[\Theta(-x) - \Theta(x)]\varphi(-x)\ ,
\\
{\cal P}_3&:& \varphi (x) 
\longrightarrow  
({\cal P}_3\varphi)(x) 
= [\Theta(x) -
\Theta(-x)]\varphi(x)\ .
\end{eqnarray}
These transformations satisfy the anti-commutation relation
%
\begin{eqnarray} 
\label{PR}
{\cal P}_i {\cal P}_j 
= \delta_{ij}+i\epsilon_{ijk} 
{\cal P}_k .
\end{eqnarray}
Since the effect of ${\cal P}_i$ on the boundary vectors $\Phi$ 
and $\Phi'$ are given by
$
\Phi 
\buildrel {\cal P}_i \over \longrightarrow \sigma_i \Phi\ , 
$
$
\Phi' 
\buildrel {\cal P}_i \over \longrightarrow \sigma_i \Phi'\ ,
$
where $\{ \sigma_i \}$ are the Pauli 
matrices,  
the transformation ${\cal P}_i$ on an element $H_U$$\in \Omega$
induces the unitary transformation 
\begin{eqnarray} 
\label{P3}
U 
\buildrel {\cal P}_i \over \longrightarrow \sigma_i U \sigma_i
\end{eqnarray}
on an element $U$ $\in$ $U(2)$.
The crucial fact is that the transformation ${\cal P}_i$ turns one 
system belonging to $\Omega$ into another 
one with {\it same spectrum}.  
In fact, with any ${\cal P}$ defined by
\begin{eqnarray} 
\label{P4}
{\cal P}
= \sum_{j = 1}^3 c_j \, {\cal P}_j
\end{eqnarray}
with real 
$c_j$ with constraint $\sum_{j = 1}^3 c_j^2 = 1$,
one has a transformation 
\begin{eqnarray} 
\label{P8}
{\cal P} H_U {\cal P} 
=H_{U_{\cal P}} 
\end{eqnarray}
where $U_{\cal P}$ is given by
%
\begin{eqnarray} 
\label{P9}
U_{\cal P} =    \sigma U \sigma
%
\ \ \
{\rm with} 
\ \ \
\sigma = \sum_{j = 1}^3 c_j\, \sigma_j .
\end{eqnarray}
One sees, from (\ref{P8}), that the system described by
the Hamiltonians $H_U$ has a family of systems
$H_{U_{\cal P}}$ 
which share the same spectrum with $H_U$. 

Let us suppose that the matrix $U$ is diagonalized with
appropriate $V \in SU(2)$ as
\begin{eqnarray} 
\label{Q10}
U = V^{-1}DV .
\end{eqnarray}
With the explicit representations
\begin{eqnarray} 
\label{Q11}
D = e^{i\xi} e^{i\rho\sigma_3}
=\left( {\matrix{{e^{i\theta _+}}&0\cr
0&{e^{i\theta _-}}\cr
}} \right) ,
\ \ \ \
\theta_\pm = \xi\pm\rho ,
\end{eqnarray}
and
\begin{eqnarray} 
\label{Q12}
V = 
e^{i{\mu \over 2}\sigma_2} e^{i{\nu \over 2}\sigma_3},
\end{eqnarray}
one can show easily that
with
$\sigma_V$ 
$=$
$e^{-i{\nu \over 2}\sigma_3}$
$e^{-i{\mu \over 2}\sigma_2}$
$e^{i{\nu \over 2}\sigma_3} \sigma_3$
$= \sigma_V^{-1}$,
one has
\begin{eqnarray} 
\label{Q13}
U=\sigma_V D \sigma_V 
\end{eqnarray}
which is of the type (\ref{P9}).
This means that $H_U$ and $H_D$ share the same spectra.
One can therefore conclude that 
the spectrum of the system described by $H_U$ is uniquely 
determined by the {\it eigenvalue} of $U$, and also 
that a point interaction
characterized by $U$ possesses the
{\rm isospectral subfamily}  
\begin{eqnarray} 
\Omega_{iso} = \left\{  H_{ V^{-1} D V } \vert 
V \in SU(2)  \right\} ,
\end{eqnarray}
which
is homeomorphic
to the 2-sphere specified by the polar angles $(\mu , \nu)$.
\begin{eqnarray} 
\Omega_{iso}
= \left\{ (\mu, \nu) \vert
\mu\in [0,\pi], \nu \in [0,2\pi) \right\}
\simeq S^2 .
\end{eqnarray}
There is of course an obvious exception to this
for the case 
of $D$ $\propto I$, in which case, $\Omega_{iso}$ 
consists only of $D$ itself.

To see the structure of the spectral space, {\it i.e.} the 
part of parameter space $U(2)$ that determines the distinct
spectrum of the system, 
it is convenient to make the spectrum 
of the system discrete.  Here, for simplicity, 
we consider
the line $x \in [-l,l]$ with Dirichlet 
boundary, $\varphi(-l)$ $= \varphi(l)$ $ = 0$.
Then, the wave function is of the form
\begin{eqnarray} 
\label{P21}
\varphi(x) 
&=& A_+ \sin{k(x-l)}  \ \ \ \ \ \ \ (x>0)
\\ \nonumber 
&=& A_- \sin{k(x+l)}  \ \ \ \ \ \ \ (x<0) .
\end{eqnarray}
%
%
%
One then has
\begin{eqnarray} 
\label{P22} 
  \left( {\matrix{{\varphi (0_+)}\cr
                  {\varphi (0_-)}\cr}
         } 
  \right) 
&=& \sin{kl} \Phi_0,
\\ \nonumber
  \left( {\matrix{{ \varphi' (0_+)}\cr
                  {-\varphi' (0_-)}\cr}
         } 
  \right) 
&=& k\cos{kl} \Phi_0,
\end{eqnarray}
with some common constant vector $\Phi_0$.
Putting this form into the connection condition (\ref{F4}),  we obtain
\begin{eqnarray} 
\label{R15}
1 + k L_0 \cot{kl} \cot{\theta_+\over 2} &=& 0,
\\ \nonumber
1 + k L_0 \cot{kl} \cot{\theta_-\over 2} &=& 0.
\end{eqnarray}
This means that the spectrum of the system is effectively 
split into that of two separate systems of same structure, 
each characterized by the parameters $\theta_+$ 
and $\theta_-$.   So the spectra of the system is
uniquely determined by two angular parameters 
$\{\theta_+, \theta_-\}$.
%
%
%
\begin{figure}[t]
\ \ \ \ \ \ 
\ \ \ \ \ \ 
\ \ \ \ \ \ 
\ \ \ \ \ \ 
\ \ \ \ \ \ 
\ \ \ \ \ \
\ \ \ \ \ \ 
\ \ \ \ \ \
\epsfxsize=11pc \epsfbox{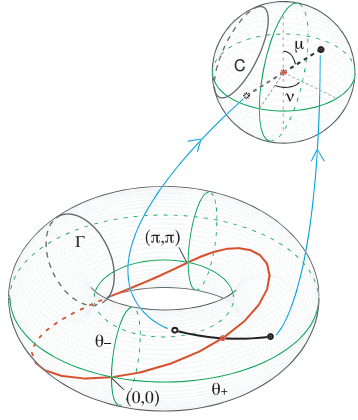}
\\
\caption{
The spectral torus
$(\theta_+, \theta_-)$
and the isospectral sphere  
$(\mu, \nu)$. 
}
\end{figure}
The entire parameter space 
$\Omega$ = $\{\theta_+, \theta_-, \mu, \nu \}$ is a product of 
spectral space  
2-torus 
\begin{eqnarray} 
\label{R20}
\Omega_{sp}
&=& \left\{ (\theta_+, \theta_-) \vert
\theta_+, \theta_-\in [0,2\pi] \right\}
\\ \nonumber
&\simeq& T^2= S^1 \times S^1 ,
\end{eqnarray}
and the isospectral space
$\Omega_{iso}$ = $\{ \mu, \nu \}$ $\simeq S^2$ (See Fig. 1).
There is another way to characterize this torus using a spin matrix
\begin{eqnarray}
\label{R21} 
\sigma_S
= \sigma_V \sigma_3\sigma_V .
\end{eqnarray}
Clearly, one has
\begin{eqnarray} 
\label{R22}
\sigma_S U \sigma_S
= U ,
\end{eqnarray}
which means that the torus $\Omega_{sp}$ is the submanifold of $\Omega$ 
that is invariant with the symmetry operation related to $\sigma_S$.
 
There is one more subtle point missing in the foregoing argument:
We note that this parameter space provides a double 
covering for the family of point inteactions $\Omega \simeq U(2)$ 
due to the arbitrariness in the 
interchange $\theta_+ \leftrightarrow \theta_-$.  
Accordingly, two systems with interchanged
values for $\theta_+$ and $\theta_-$ are isospectral.
So the space of distinct spectra $\Sigma$ 
is the torus 
$T^2 = \{ (\theta_+,\theta-) | \theta_\pm \in [0,2\pi)\}$
subject to the identification    
$(\theta_+,\theta_-)$ $\equiv (\theta_-,\theta_+)$.
Thus we have
\begin{eqnarray} 
\label{R30} 
\Sigma = \{ Spec(H_U) | U \in \Omega \} = T^2/{\bf Z}_2 , 
\end{eqnarray}
which is homeomorphic to a M{\"o}bius strip with boundary \cite{TF01}.

Looking this double covering nature of the spectral torus from the other side,
one may also say that on the isospectral $S^2$,  the point interactions 
corresponding to the two polar opposite positions occupy a special positions, 
because they belong to the same spectral $T^2$ sharing 
the same symmetry invariance (\ref{R22}).  We call these pairs {\it dual} to each 
other.  
One particular example is this duality is  is given by 
$\{ \mu = \pi/2, \nu=0\}$ and $\{ \mu = \pi/2, \nu=\pi\}$.  These pairs belong to
the parity (in original left-right sense) invariant torus $\sigma_1 U \sigma_1$ $= U$.
One can check immediately that this essentially is the duality between the
$\delta$ interaction system and $\varepsilon$ interaction system 
with opposite parity states which has appeared in the previous section.
%
%
%
\begin{figure}[t]
\ \ \ \ \ \ 
\ \ \ \ \ \ 
\epsfxsize=9pc \epsfbox{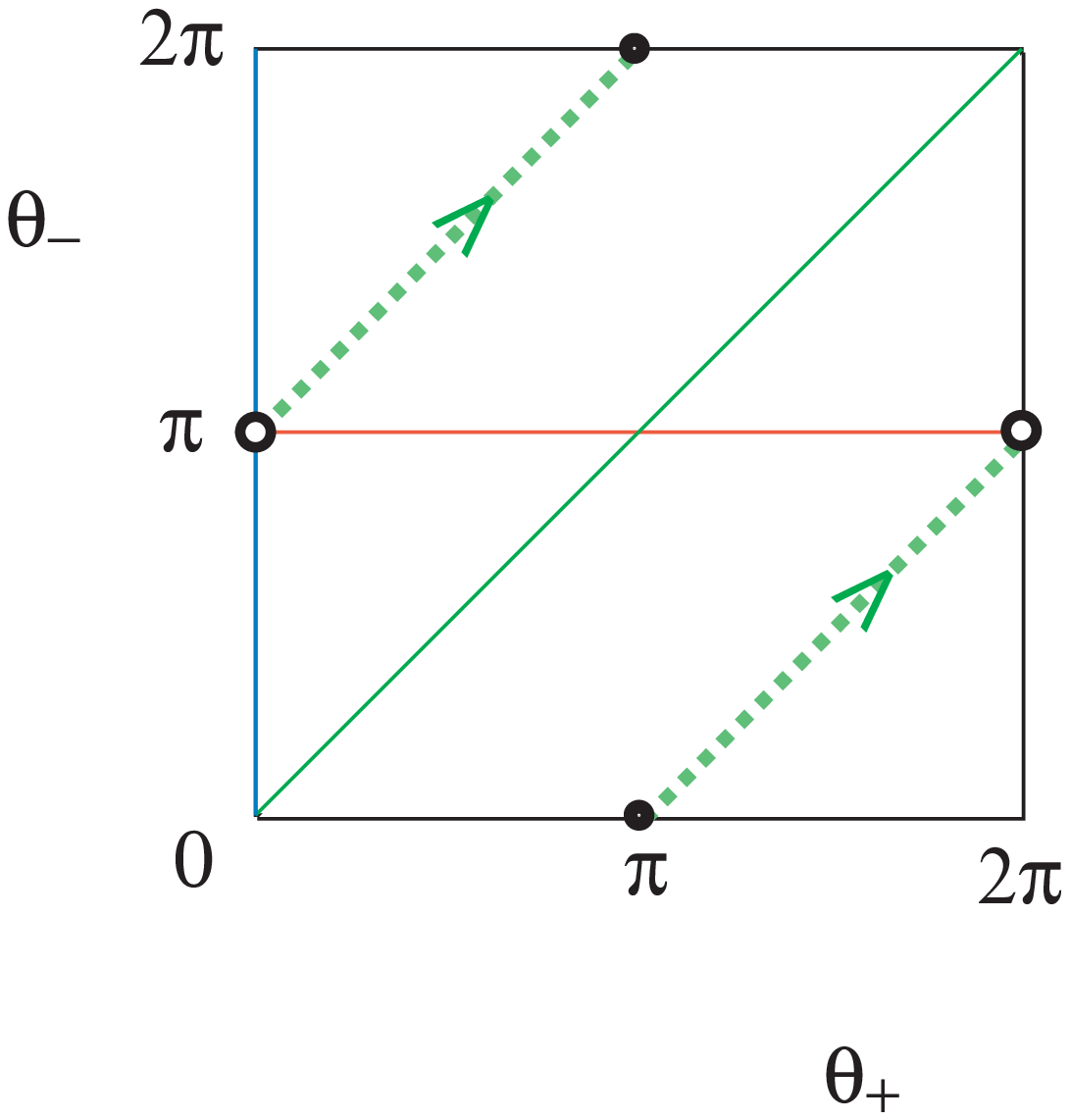}
\ \ \ \ \ \
\epsfxsize=12pc \epsfbox{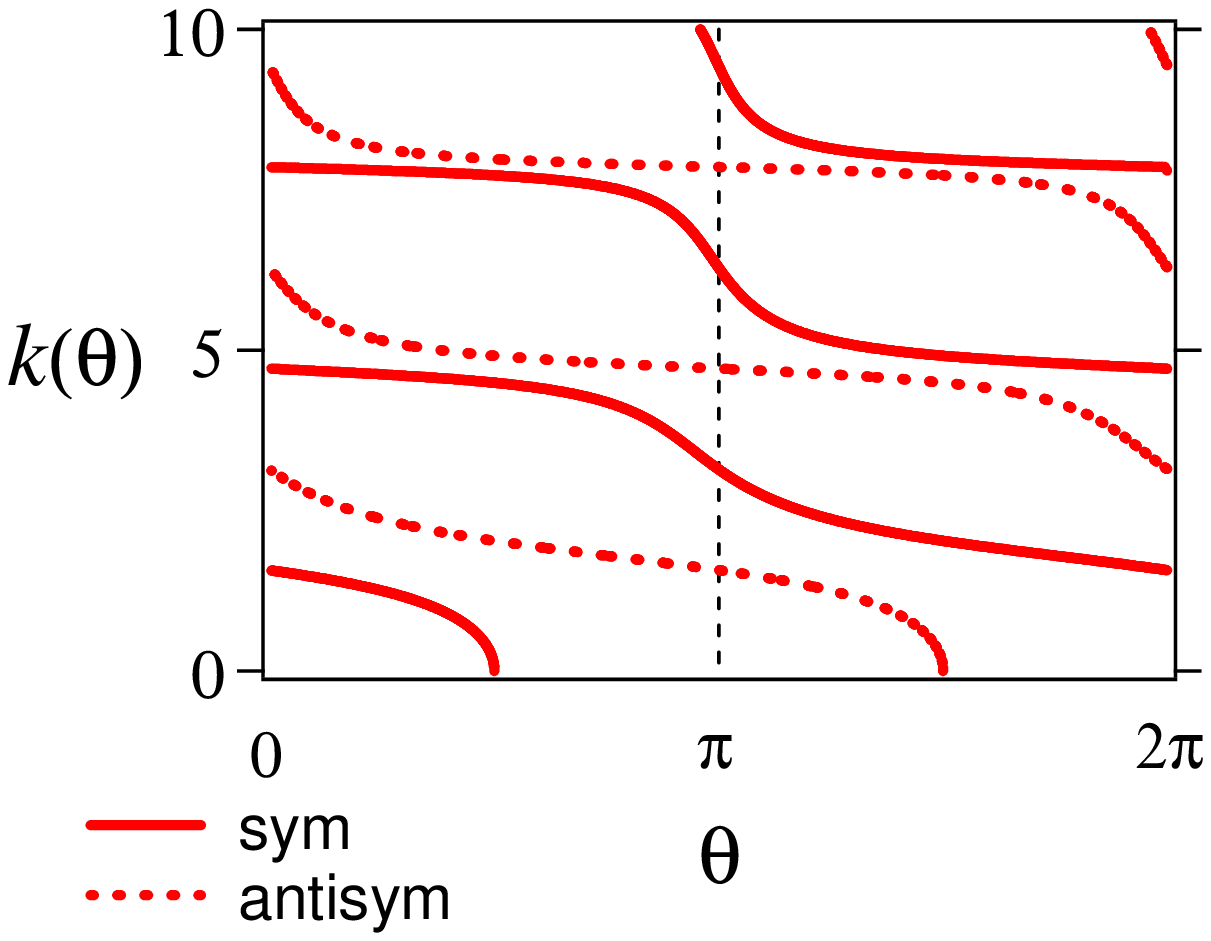}
\caption{
Spiral anholonomy on the torus $\{\theta_+. \theta_-\}$.
}
\end{figure}

An intriguing phenomenon is revealed by a closer examination of the
spectral equation (\ref{R15}).   Obviously, energy spectrum as a function of
the parameter $\theta_+$ or  $\theta_-$ has to be a $2\pi$-periodic function.
With elementary calculation, however, one can explicitly
see the relations
\begin{eqnarray} 
\label{R31} 
{{dk\ } \over {d\theta_+}} ,
\ \ 
{{dk\ } \over {d\theta_-}} < 0. 
\end{eqnarray}
The only way to reconcile these two facts is through the ``spectral flow'' ;
namely, when $\theta_\pm$ is increased by $2\pi$, an energy eigenstate is shifted
to a lower eigenstate while the spectra as a whole are unchanged \cite{CH98}.  
The situation becomes clear by the illustration shown in Fig. 2, where the spectra is
plotted as functions of $\theta =$$\theta_+ =$ $\theta_- + \pi/2$.  
The root of this phenomenon is the nontrivial topology of the spectral space $T^2$,
as expressed in the homotopy group $\pi_1(T^2)={\bf Z} \times {\bf Z}$.
This type of ``spiral anholonomy'' has been known in quantum physics only 
in non-Abelian gauge theories until now.

At this point, some readers might be wondering whether nontrivial topology of
the isospectral sphere, $\pi_2(S^2)={\bf Z}$, has any observable consequences.
We simply note that affirmative answers are given in the form of 
Berry phase \cite{CS96,BP98}. 
%
\section{A Prospect}
Immediate and useful extensions of our treatment exist for the quantum mechanics
on the graphs \cite{EX96}.  The analysis of so-called ``X-junction'' 
in terms of  $U(4)$ parameter space appears to have particular urgency because of
its potential relevance to the quantum informational devices \cite{BH02}.     

\medskip
This work has been supported in part by 
the Monbu-Kagakusho Grant-in-Aid for Scientific Research 
(No. (C)13640413).



\begin{thebibliography}{99}
%
\bibitem{FT00}
T. F{\"u}l{\"o}p and I. Tsutsui
Phys. Lett.
{\bf A264}, 366 (2000).
%
\bibitem{SE86}
P. {\v S}eba, 
Czech. J. Phys. {\bf B36}, 667 (1986). 
%
\bibitem{AG88}
S. Albeverio, F. Gesztesy, R. H\o egh-Krohn and H. Holden, 
{\em Solvable Models in Quantum Mechanics} 
(Springer, Heidelberg, 1988).
%
\bibitem{CS98}
T. Cheon and T. Shigehara, 
Phys. Lett. {\bf A243}, 111 (1998).
%
\bibitem{EN01}
P. Exner, H. Neidhardt and A. Zagrebnov, 
Comm. Math. Phys. {\bf 224}, 593 (2001).
%
\bibitem{CS99}
T. Cheon and T. Shigehara, 
Phys. Rev. Lett. {\bf 82},  2539 (1999).
%
\bibitem{TF00}
I. Tsutsui, T. F{\"u}l{\"o}p and T. Cheon,
J. Phys. Soc. Jpn,
{\bf 69}, 3473 (2000).
%
\bibitem{CF01}
T. Cheon, T. F{\"u}l{\"o}p and I. Tsutsui,
Ann. of Phys. (NY) {\bf 294}, 1 (2001).
%
\bibitem{TF01}
I. Tsutsui, T. F{\"u}l{\"o}p and T. Cheon,
J. Math. Phys. {\bf 42} 5687 (2001).
%
\bibitem{CH98}
T. Cheon, 
Phys. Lett. {\bf A248}, 285 (1998). 
%
\bibitem{CS96}
T. Cheon and T. Shigehara, 
Phys. Rev. Lett. {\bf 76},  1770 (1996).
%
\bibitem{BP98}
M. Brazovskaia and P. Pieranski,
Phys. Rev. {\bf E58},  4076 (1998).
%
\bibitem{EX96}
P. Exner,
Lett. Math. Phys. {\bf 38},  313 (1996); 
{\it ibid.} {\bf 42} 193 (1997) .
%
\bibitem{BH02}
S. Bose and D. Home, 
Phys. Rev. Lett. {\bf 88},  050401 (2002).
%
%
\end{thebibliography}
\end{document}